\documentclass[aps,prl,showpacs,twocolumn,superscriptaddress]{revtex4}
\usepackage{bm,color}
\usepackage{graphicx}
\usepackage{amsmath}

\begin{document}
\title {Spin model of magnetostrictions in multiferroic Mn perovskites}

\author{Masahito Mochizuki}
\affiliation{Department of Applied Physics, The University of Tokyo,
7-3-1, Hongo, Bunkyo-ku, Tokyo 113-8656, Japan}

\author{Nobuo Furukawa}
\affiliation{Department of Physics, Aoyama Gakuin University,
Fuchinobe 5-10-1, Sagamihara, 229-8558 Japan}
\affiliation{Multiferroics Project, ERATO, Japan Science and Technology 
Agency (JST) 
c/o Department of Applied Physics, The University of Tokyo, 
Tokyo 113-8656, Japan}

\author{Naoto Nagaosa}
\affiliation{Department of Applied Physics, The University of Tokyo,
7-3-1, Hongo, Bunkyo-ku, Tokyo 113-8656, Japan}
\affiliation{Cross-Correlated Materials Research Group (CMRG)
and Correlated Electron Research Group (CERG), 
RIKEN-ASI, Saitama 351-0198, Japan}

\begin{abstract}
We theoretically study origins of the ferroelectricity in the 
multiferroic phases of the rare-earth ($R$) Mn perovskites, $R$MnO$_3$, 
by constructing a realistic spin model including the spin-phonon coupling, 
which reproduces the entire experimental phase diagram in the plane of 
temperature and Mn-O-Mn bond angle for the first time. 
Surprisingly we reveal a significant contribution of the symmetric 
($\bm S \cdot \bm S$)-type magnetostriction to the ferroelectricity
even in a spin-spiral-based multiferroic phase, which can be larger than 
the usually expected antisymmetric ($\bm S \times \bm S$)-type contribution.
This explains well the nontrivial behavior of the electric polarization.
We also predict the noncollinear deformation of the $E$-type spin 
structure and a wide coexisting regime of the $E$ and 
spiral states, which resolve several experimental puzzles.
\end{abstract}
\maketitle
Frustrating spins in magnets often exhibit not only nontrivial 
orders but also intriguing switching and dynamical 
phenomena associated with phase competitions. 
Effective reduction of the spin-exchange energy due to the 
frustration increases the
relative importance of other tiny interactions such as the
Dzyaloshinskii-Moriya (DM) interaction, single-ion anisotropy, 
and spin-phonon coupling.
Their fine energy balance results in keen conflict of various states.
This enables us to achieve sensitive phase controls and huge 
responses leading to new functionalities of materials, and also
provides challenging issues for fundamental science.

The rare-earth ($R$) Mn perovskites, $R$MnO$_3$, offer one of the
most typical examples.
In this class of materials, the nearest-neighbor spin exchange 
is very small ($\sim$1 meV) relative to that in 
other perovskite compounds (e.g. $\sim$15 meV in LaTiO$_3$)
due to the cancellation of exchange contributions
from $t_{2g}$ and $e_g$ orbital sectors~\cite{Mochizuki09}.
Consequently the next-neighbor antiferromagnetic (AFM) coupling becomes
comparable to the nearest-neighbor ferromagnetic (FM) coupling.
Their frustration gives rise to various competing phases
including multiferroic phases where the frustration-induced 
nontrivial spin order generates ferroelectric 
polarization $\bm P$~\cite{Kimura03a,ReviewMF}.

Recent experiments revealed spectacular magnetoelectric (ME) phenomena
in these multiferroic phases, i.e., 
magnetic-field-induced $\bm P$ flops~\cite{Kimura03a,Kimura05}, 
colossal magnetocapacitance~\cite{Kimura05,Kagawa09,SchrettleCD08}, 
and electromagnons~\cite{Pimenov06a,Kida09}.
To study and/or control these cross-correlation phenomena, 
thorough understanding of the magnetic structures, the phase competitions, 
and the coupling between magnetic and ferroelectric orders based on a 
reliable model is essential. 

However, there still remain many experimental observations, which 
are not understood theoretically.\\
(i) Multifurcation of the sinusoidal collinear phase at higher
temperature ($T$) into four low-$T$ phases depending on the ionic
$R$-site radius ($r_R$)~\cite{Ishiwata10}.\\
(ii) Nontrivial $r_R$-dependence of the magnitude and direction
of $\bm P$~\cite{Ishiwata10}.\\
(iii) Apparently contradicting neutron-scattering results on the
magnetic commensurability in the compounds with small 
$r_R$~\cite{Munoz01,Brinks01,Munoz02,YeF07}.\\
(iv) Anomalous $T$-dependence of $|\bm P|$ for YMnO$_3$ and 
ErMnO$_3$~\cite{Ishiwata10}.

In this Letter, we study theoretically the interplay of symmetric 
($\bm S \cdot \bm S$)-type magnetostriction (MS) and antisymmetric 
($\bm S \times \bm S$)-type MS in a realistic spin model for $R$MnO$_3$, 
and resolve all the puzzles listed above. 
We find a large ($\bm S \cdot \bm S$) contribution
to the ferroelectricity even in a spiral spin phase. 
This mechanism is generally expected in all the spin-spiral-based 
multiferroics.
\begin{figure}[tdp]
\includegraphics[scale=1.0]{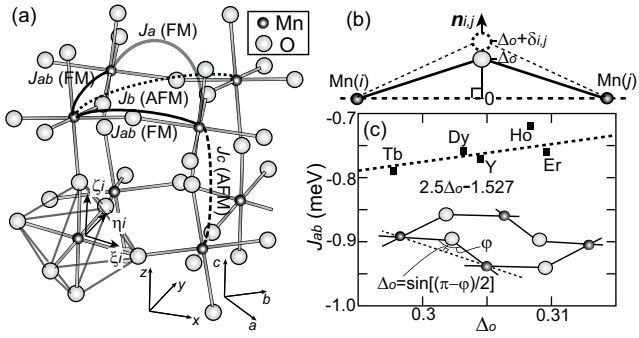}
\caption{(a) Spin exchanges in $R$MnO$_3$ 
where FM (AFM) denotes (anti)ferromagnetic exchange. 
(b) Mn($i$)-O-Mn($j$) bond in the orthorhombic lattice 
and local vector $\bm n_{i,j}$. The O ion is displaced from its 
cubic position (0) to the orthorhombic position ($\Delta_o$) at higher 
$T$. A further shift $\delta_{i,j}$ along $\bm n_{i,j}$ can be induced 
at low $T$ via the spin-lattice coupling. 
(c) $\Delta_o$ vs $J_{ab}$ for several $R$ species
calculated in Ref.~\cite{Mochizuki09}.
Here $\Delta_o$ is normalized by the MnO bond length.}
\label{Fig01}
\end{figure}
We start with a model in which the Mn $S$=2 spins are treated as classical 
vectors on a cubic lattice. 
A similar model has been examined in Refs.~\cite{Mochizuki09,MochizukiCD10} 
which gives the transition between two types of multiferroic spiral spin 
phases and explains several ME phenomena quantitatively.
Here we further include the lattice degrees of freedom. 
This enables us to study the effect of the ($\bm S \cdot \bm S$)-type 
MS as a source of the above puzzles, 
which has been missed thus far.

The Hamiltonian is given by
\begin{eqnarray}
\mathcal{H}&=&
\sum_{<i,j>} J_{ij} \bm S_i \cdot \bm S_j
+D \sum_{i} S_{\zeta i}^2  \nonumber \\
&+&E\sum_{i}(-1)^{i_x+i_y}(S_{\xi i}^2-S_{\eta i}^2) \nonumber \\
&+&\sum_{<i,j>}\bm d_{ij}\cdot(\bm S_i \times \bm S_j)
+K \sum_i(\delta_{i,i+\hat x}^2+\delta_{i,i+\hat y}^2),
\end{eqnarray}
where $i_x$, $i_y$, $i_z$ represent the integer coordinates of 
the $i$th Mn ion with respect to the cubic $x$, $y$ and $z$ axes. 

The first term describes the spin-exchange 
interactions as shown in Fig.~\ref{Fig01}(a).
The second and third terms stand for the single-ion anisotropy. 
For the local axes $\bm \xi_i$, $\bm \eta_i$ and $\bm \zeta_i$ 
attached to each MnO$_6$ octahedron, we use the structural 
data of DyMnO$_3$~\cite{Alonso00}. 
The fourth term denotes the DM interaction. 
The DM vectors $\bm d_{ij}$ 
are expressed using five DM parameters, $\alpha_{ab}$, $\beta_{ab}$,
$\gamma_{ab}$, $\alpha_c$ and $\beta_c$, as given in Ref.~\cite{Solovyev96}
because of the crystal symmetry.
The last term represents 
the lattice elastic term with $K$ being the elastic constant.
Here $\delta_{i,j}$ is a shift of the O ion between 
$i$th and $j$th Mn ions normalized by the MnO bond length. 
Note that the O ion in the orthorhombic lattice is already displaced from 
its cubic position. 
We consider $\delta_{i,j}$ as a further shift from the 
position at higher $T$ in the presence of magnetic order at low $T$. 

Since the nearest-neighbor FM coupling in $R$MnO$_3$ 
is sensitive to the Mn-O-Mn bond angle, 
we consider the Peierls-type spin-phonon coupling 
$J_{ij}=J_{ab}+J_{ab}'\delta_{i,j}$
for the in-plane Mn-O-Mn bonds
where $J_{ab}'$=$\partial J_{ab}$/$\partial \delta$~\cite{Kaplan09}.
We assume that the shift of O ion $\delta_{i,j}$ occurs 
along the local axis $\bm n_{i,j}$ directing from its cubic position (0)
to the orthorhombic position ($\Delta_o$) at higher $T$ [see
Fig.~\ref{Fig01}(b)]. Then the positive (negative) shift decreases 
(increases) the Mn-O-Mn bond angle.

The values of $J_{ab}$, $J_c$, $J_b$, $D$, $E$, and five DM parameters 
have been microscopically determined in Ref.~\cite{Mochizuki09} 
for several $R$MnO$_3$ compounds.
Except for $J_b$, they are nearly invariant upon the $R$-site variation 
in the vicinity of the multiferroic phases. 
We fix $J_{ab}$=$-$0.8, $J_c$=1.25, $D$=0.2, $E$=0.25, 
($\alpha_{ab}$, $\beta_{ab}$, $\gamma_{ab}$)=(0.1, 0.1, 0.14), and 
($\alpha_c$, $\beta_c$)=(0.42, 0.1). 
Here the energy unit is meV. 
We also find that very weak FM exchange $J_a$ is necessary to produce 
the $E$ phase, and adopt $J_a$=$-$0.1. The value of $K$ is chosen to
be 500 so as to reproduce the experimental $P$ in the $E$ phase
[see Fig.~\ref{Fig03}(a)].
We obtain the value of $J_{ab}'$ from the $\Delta_o$ dependence of 
$J_{ab}$ for several $R$ species [see Fig.~\ref{Fig01}(c)], 
which gives 
$J_{ab}'$=$\partial J_{ab}$/$\partial \Delta_o$=2.5.

We treat $J_b$ as a variable which increases (decreases) as $r_R$ 
decreases (increases). 
This is because the exchange path for $J_b$ contains two O $2p$ 
orbitals, and the orthorhombic distortion, whose magnitude is 
controlled by $r_R$, enhances their hybridization.
We find that overall features of the phase evolution upon the 
$R$-site variation are reproduced as a function of $J_b$ even without
considering the slight $R$-dependence of other parameters.

We analyze the above model using the replica exchange Monte-Carlo 
(MC) method~\cite{Hukushima96}. Both spins and oxygen positions are
updated in the simulation, and each exchange sampling is taken after 
400 standard MC steps. 
Typically, we perform 600 exchanges for a system of $N$=
48$\times$48$\times$6 sites along $x$, $y$ and $z$ axes
with periodic boundaries. 
We identify transition points and spin structures from $T$ profiles 
of specific heat and spin-helicity vector 
$\bm h=\frac{1}{2N}\sum_{i} 
(\bm S_i \times \bm S_{i+\hat x} 
+ \bm S_i \times \bm S_{i+\hat y})/S^2$.
We also calculate spin correlations in the momentum space
by the Fourier transformation of spin configurations.

\begin{figure}[tdp]
\includegraphics[scale=1.0]{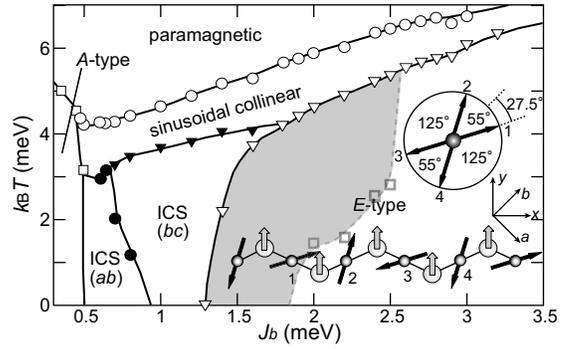}
\caption{Theoretical phase diagram of $R$MnO$_3$ in the plane of 
$J_b$ and $T$. Here ICS denotes the incommensurate spiral phase.
In the shaded area, the $E$ and ICS states can coexist. 
Inset shows real-space spin configuration
of the $E$ phase. Shifts of the O ions due to the 
($\bm S \cdot \bm S$)-type magnetostriction are shown by gray arrows.}
\label{Fig02}
\end{figure}
We first display the theoretical $T$-$J_b$ phase diagram 
in Fig.~\ref{Fig02}, which successfully reproduces the experimentally
observed phase evolutions~\cite{Tokura06}.
More concretely, the following four phases
successively emerge at low $T$ as $J_b$ decreases: the $A$, $ab$-spiral, 
$bc$-spiral, and $E$ phases. 
In the $A$ ($E$) phase, the FM (up-up-down-down) Mn-spin
layers stack antiferromagnetically, while in the $ab$ ($bc$) spiral
phase, the Mn spins rotate within the $ab$ ($bc$) plane ($Pbnm$
setting) to form transverse cycloids~\cite{Kenzelmann05,Yamasaki08}.
As $T$ decreases, these four phases emerge with multifurcation 
from the sinusoidal collinear phase at higher $T$ where 
the collinear Mn spins are sinusoidally
modulated in amplitude.
The spin structure is commensurate (C) with $q_b$=0.5 in the 
$E$ phase, whereas it is incommensurate (IC) in the $ab$ and $bc$ 
spiral phases. 
Importantly, the sinusoidal collinear state is also IC 
even above the $E$ phase (e.g. $q_b$=0.458 for $J_b$=2.4), and 
the spin-phonon coupling is a source of the IC-C transition 
with lowering $T$.

In the $ab$ ($bc$) spiral phase, it has been naively believed that the 
antisymmetric ($\bm S \times \bm S$)-type MS 
induces the ferroelectric polarization $\bm P$$\parallel$$a$ 
($\bm P$$\parallel$$c$)~\cite{Katsura05,Sergienko06a,Mostovoy06}.
However, the observed $P$ in the $ab$ spiral phase is much larger than 
that in the $bc$ spiral phase. 
For instance, the $P_a$ in the $ab$ spiral phase of DyMnO$_3$ under 
$\bm H$$\parallel$$b$ is 2.5 times larger than $P_c$ in the $bc$ 
spiral phase at $H$=0~\cite{Kimura05,Kagawa09}.
Moreover, in Eu$_{0.6}$Y$_{0.4}$MnO$_3$, the $P_a$ at 
$H$=0 is approximately 10 times larger than $P_c$
under $\bm H$$\parallel$$a$, which excludes the influence of $f$
moments as its origin because of their absence~\cite{Yamasaki07b}.
This is quite puzzling since we expect nearly identical strength of the 
($\bm S \times \bm S$)-type MS in these two phases.
Recent first-principles study also suggested different mechanisms of $P$
between the two spiral phases~\cite{Malash08}.
To solve this issue, we calculate the polarization due to the 
($\bm S \cdot \bm S$)-type MS,
$\bm P_{\rm S}$=($\tilde P_a$, $\tilde P_b$, $\tilde P_c$)
from the oxygen shifts. 
Because of the staggered local axes $\bm n_{i,j}$ 
on the zigzag Mn-O chain, $\tilde P_\gamma$ 
($\gamma$=$a$, $b$, $c$) is given by 
\begin{equation}
\tilde P_\gamma=-\frac{\Pi_\gamma}{N}
\sum_{i}[(-1)^{i_x+i_y+m} 
\delta_{i,i+\hat x} + (-1)^{i_x+i_y+n} \delta_{i,i+\hat y}], \nonumber
\end{equation}
where ($m$, $n$)=(0, 0) for $\gamma$=$a$, 
($m$, $n$)=(1, 0) for $\gamma$=$b$, and ($m$, $n$)=($i_z$+1, $i_z$+1) 
for $\gamma$=$c$. 
Here the constant $\Pi_\gamma$ is calculated to be 
4.6$\times$10$^5$ $\mu C$/$m^2$ for $\gamma$=$a$ and $b$, 
and 3.3$\times$10$^5$ $\mu C$/$m^2$ for $\gamma$=$c$ 
from lattice parameters using the point-charge model. 

\begin{figure}[tdp]
\includegraphics[scale=1.0]{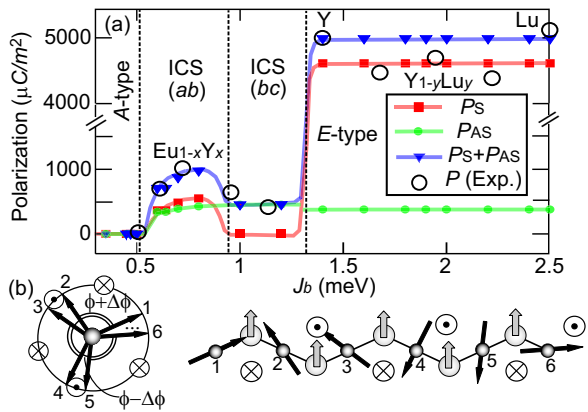}
\caption{(Color online) (a) Polarizations vs $J_b$ at $T$$\rightarrow$0, 
i.e., ($\bm S \cdot \bm S$) contribution $P_{\rm S}$, 
($\bm S \times \bm S$) contribution $P_{\rm AS}$, 
and experimentally measured $P$ in 
Eu$_{1-x}$Y$_x$MnO$_3$ and Y$_{1-y}$Lu$_y$MnO$_3$~\cite{Ishiwata10}.
The summation $P_{\rm S}$+$P_{\rm AS}$ reproduces
the experimental $P$ well.
(b) Alternation of the spin angles in the $ab$ spiral state due to the 
staggered DM vectors is illustrated in an exaggerated manner 
where $\odot$ ($\otimes$) denotes the positive (negative) 
$c$-component of the vector. Shifts of the O ions due to the 
($\bm S \cdot \bm S$)-type magnetostriction are shown by gray arrows.}
\label{Fig03}
\end{figure}
In Fig.~\ref{Fig03}(a), we plot calculated $P_{\rm S}$ at 
$T$$\rightarrow$0 as functions of $J_b$.
Surprisingly we find a finite $P_{\rm S}$ in the $ab$ spiral phase 
(e.g., $P_{\rm S}$$\sim$500 $\mu C$/$m^2$ for $J_b$=0.7), 
while it is zero in the $bc$ spiral phase. 
This can be understood as follows [see also Fig.~\ref{Fig03}(b)]. 
On the in-plane chains, the $c$-axis components of the 
DM vectors are arranged in the staggered way. 
As a result, the spin rotation angles in the $ab$ spiral 
become subject to an alternate modulation~\cite{Note1}. 
Then the O ions between two spins with a smaller angle of 
$\phi$$-$$\Delta\phi$ (a larger angle of $\phi$$+$$\Delta\phi$) shift 
negatively (positively) to strengthen (weaken) the FM exchange 
through increasing (decreasing) the Mn-O-Mn bond angle.
These shifts generate a uniform component
resulting in the ferroelectric polarization.
In fact, the spin rotation angles in the $bc$ spiral are also 
subject to the alternate modulation because of the staggered
$a$-axis components of the DM vectors. However the induced O shifts 
are opposite between neighboring $ab$ planes, which results in their 
perfect cancellation.

We also show $J_b$-dependence of $P_{\rm AS}$ of ($\bm S \times \bm S$) 
origin at $T$$\rightarrow$0 in Fig.~\ref{Fig03}(a). Since the $P_{\rm AS}$ 
consists of two contributions, i.e. the electronic and
the lattice-mediated ones~\cite{Malash08} and the former one is difficult 
to evaluate by the spin model, we calculate $P_{\rm AS}$ from the 
spin helicity $\bm h$. Note that $P_{\rm AS}$ is proportional 
to $|\bm h|$ and the observed $P$ in the $bc$ spiral phase is
purely of ($\bm S \times \bm S$) origin.
In addition, we plot the experimentally measured $P$ 
of Eu$_{1-x}$Y$_x$MnO$_3$ and Y$_{1-y}$Lu$_y$MnO$_3$
for comparison~\cite{Ishiwata10}, whose $P$ originates purely 
from the Mn-spin order because of the absence of $f$ 
moments~\cite{Prokhnenko07}. Effective $r_R$ and $J_b$ of these
solid solutions are evaluated by interpolations.

We find that the summation
$P_{\rm S}$+$P_{\rm AS}$ reproduces well the experimental $P$.
Here we emphasize that only the elastic constant $K$ is an uncertain
parameter in our model, and once we determine its value so as to 
reproduce the experimental $P$ in the $E$ phase, the behaviors of $P$ 
in the spiral phases are reproduced almost perfectly.
Moreover it turns out that the ($\bm S \cdot \bm S$) contribution 
$P_{\rm S}$ can be comparable to or even larger than the 
($\bm S \times \bm S$) contribution $P_{\rm AS}$ in the $ab$ spiral phase. 
This explains why $P$ in the $ab$ spiral phase is much larger than 
that in the $bc$ spiral phase. 

Next we discuss the $E$ phase. 
Interestingly we find a finite $c$ component of the spin-helicity vector 
$\bm h$ in this phase, indicating that its spin structure is 
not collinear in contrast to what we have believed so far, 
but its up-up-down-down structure 
is subject to a spiral modulation within the $ab$ plane. 
The inset of Fig.~\ref{Fig02} illustrates the real-space spin 
configuration of the $E$ phase, which indeed shows the 
elliptically deformed $ab$-plane cycloid. 
Calculating the $T$-dependence of the expectation value for each 
term in the Hamiltonian, we find that the single-ion anisotropy or 
alternation of the in-plane easy magnetization axes due to the 
$d_{3x^2-r^2}$/$d_{3y^2-r^2}$-type orbital ordering is an origin of
the cycloidal deformation. 
This predicted deformation should be confirmed in a future polarized 
neutron-scattering experiment.

With dominant up-up-down-down spin $b$-axis components,
the O ions between nearly (anti)parallel Mn-spin pairs shift
negatively (positively) to modulate the FM exchanges, 
which results in the ferroelectric polarization 
(see inset of Fig.~\ref{Fig02})~\cite{Sergienko06b,Picozzi07}.
In Fig.~\ref{Fig03}(a), we indeed see a very large $P_{\rm S}$ 
($\sim$4600 $\mu C$/$m^2$) in the $E$ phase. 
We also expect a small but finite ($\bm S \times \bm S$) contribution 
$P_{\rm AS}$ due to the cycloidal deformation.
The calculated $P_{\rm S}$, $P_{\rm AS}$ and their sum 
$P_{\rm S}$+$P_{\rm AS}$ in the $E$ phase are invariant upon the 
$J_b$ variation in agreement with the experiment~\cite{Ishiwata10}.

Finally we discuss the coexistence of the $E$ state and 
the IC-spiral (ICS) state. 
In the shaded area of Fig~\ref{Fig02}, we obtain 
coexisting solutions when we perform the MC calculation starting 
from the ICS state as an initial configuration. 
The spin-phonon coupling or the ($\bm S \cdot \bm S$)-type 
MS make the transition between ICS and $E$ phases
of strong first order. As a result, 
although the energy comparison gives the transition line 
between them as indicated by the solid line, 
the ICS state has a deep energy minimum even in the $E$ phase. 
This can result in their coexistence.
Such a coexistence easily occurs in reality since the system enters 
into the $E$ phase necessarily via the IC sinusoidal 
collinear phase with lowering $T$.

\begin{figure}[tdp]
\includegraphics[scale=1.0]{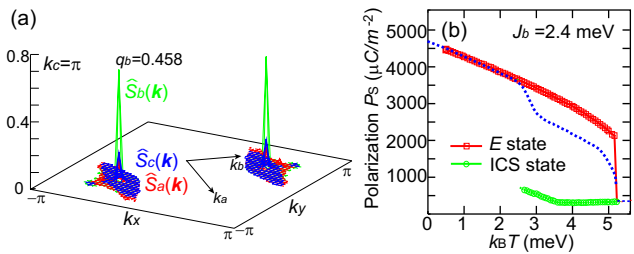}
\caption{(Color online) (a) Calculated spin-correlation functions 
in the momentum space when the $E$ and incommensurate spiral (ICS) 
states coexist. 
Here $\hat S_\gamma(\bm k)$ denotes the correlation function for the 
spin $\gamma$-axis components. 
(b) Calculated $T$ dependence of $P_{\rm S}$ in the pure $E$ state and 
that in the ICS state for 2.6$<k_{\rm B}T<$5.1. 
Since the $P_{\rm S}$ in the coexisting regime necessarily becomes 
smaller than that for the pure $E$ state, an anomaly can be observed 
in the $T$ profile of $P$ as indicated by the dashed line.}
\label{Fig04}
\end{figure}
To understand the contradicting neutron-scattering 
results~\cite{Munoz01,Brinks01,Munoz02,YeF07},
we calculate the spin-correlation functions in the momentum space 
for a coexisting solution obtained in the MC calculation.
We find only the spin $b$-axis component 
has sharp peaks at $q_b$=$\pm$0.458, while the other components have very
small peaks as shown in Fig.~\ref{Fig04}(a).
This seems as if the spin structure were IC {\it collinear}, 
which leads to the confusion. 
Observations of the IC wave numbers $q_b$$\sim$0.43 in $R$MnO$_3$ with 
$R$=Ho~\cite{Brinks01} and Er~\cite{YeF07} as well as 
a report of the IC collinear state in YMnO$_3$~\cite{Munoz02} can be 
attributed to this coexisting state, while 
a report of the commensurate $q_b$=0.5 in HoMnO$_3$~\cite{Munoz01} is to 
the pure $E$ state.

When 1.8$<J_b<$2.5, the energy minimum of the ICS state 
disappears as $T$ decreases. 
In this case, an anomaly should appears in the $T$ 
profile of $P$ or dielectric constant [see Fig.~\ref{Fig04}(b)].
A recent experiment indeed observed such an anomaly in YMnO$_3$ and 
ErMnO$_3$~\cite{Ishiwata10}, which strongly evidences the coexistence. 
The coexistence together with the ($\bm S \cdot \bm S$) contribution
in the $E$ phase should be seriously considered also 
when we interpret the experimental results for $R$MnO$_3$ with 
$R$=Y, Ho, ...,Lu, like the strange electromagnon spectra in the optical 
spectroscopy~\cite{Takahashi10}.

In summary, we have theoretically studied the magnetic structures 
and the ME coupling in $R$MnO$_3$ by using
a realistic spin model including the spin-phonon coupling. 
We have succeeded in reproducing the entire phase diagram of $R$MnO$_3$
for the first time, and have revealed the cooperative contributions of
symmetric ($\bm S \cdot \bm S$)-type and antisymmetric 
($\bm S \times \bm S$)-type MSs to the ferroelectricity 
in the $ab$ spiral phase.
This mechanism is generic and is relevant to all the spin-spiral
multiferroics.
We have also found the cycloidal spin deformation in the $E$ phase, 
and the coexistence of the $E$ and ICS states.
On these basis, the nontrivial behavior of $\bm P$
and several puzzles in the experiments have been explained. 
Our model gives a firm basis for studying and controlling 
the intriguing cross-correlation phenomena in $R$MnO$_3$.

The authors are grateful to Y. Tokura, S. Ishiwata, F. Kagawa, 
D. Okuyama, and T. Arima for discussions. This work was supported 
by Grant-in-Aids 
(22740214, 21244053, 17105002, 19048015, and 19048008), G-COE Program
(``Physical Sciences Frontier") and
NAREGI Project from MEXT of Japan, and
Funding Program for World-Leading Innovative R$\&$D on Science 
and Technology (FIRST Program) from JSPS.


\end{document}